\documentclass[aps,prd,twocolumn,showpacs,preprintnumbers,superscriptaddress]{revtex4-1}  
\usepackage{graphicx}  
\usepackage{dcolumn}   
\usepackage{bm}        
\usepackage{amssymb}   
\usepackage{comment}
\usepackage{amsmath}
\usepackage{amsfonts}
\usepackage{slashed}
\usepackage{soul}
\usepackage{graphicx}
\usepackage{float}
\usepackage{cancel}
\usepackage{color}
\usepackage{cancel}
\newcommand{\beq}{\begin{equation}}
\newcommand{\eeq}{\end{equation}}
\newcommand{\beqa}{\begin{eqnarray}}
\newcommand{\eeqa}{\end{eqnarray}}
\newcommand{\bit}{\begin{itemize}}
\newcommand{\eit}{\end{itemize}}
\newcommand*{\mathcolor}{}
\def\mathcolor#1#{\mathcoloraux{#1}}
\newcommand*{\mathcoloraux}[3]{%
  \protect\leavevmode
  \begingroup
    \color#1{#2}#3%
  \endgroup
}
\begin{document}

\widetext
\leftline{To be submitted to PRD}
\preprint{KCL-PH-TH/2015-04}


\title{
$SO(10)$ Grand Unification in $M$ theory on a $G_2$ manifold}

\author{Bobby S. Acharya}
\email{bobby.acharya@kcl.ac.uk}
\affiliation{International Centre for Theoretical Physics, I-­34151 Trieste, ITALY}
\affiliation{Department of Physics, King's College London, London, WC2R 2LS, UK}
\author{Krzysztof Bo\.{z}ek}
\email{krzysztof.bozek@kcl.ac.uk}
\affiliation{Department of Physics, King's College London, London, WC2R 2LS, UK}
\author{Miguel Crispim Rom\~{a}o}
\email{m.crispim-romao@soton.ac.uk}
\affiliation{School of Physics and Astronomy, University of Southampton, Southampton, SO17 1BJ, UK}
\author{Stephen F. King}
\email{s.f.king@soton.ac.uk}
\affiliation{School of Physics and Astronomy, University of Southampton, Southampton, SO17 1BJ, UK}
\author{Chakrit Pongkitivanichkul}
\email{chakrit.pongkitivanichkul@kcl.ac.uk}
\affiliation{Department of Physics, King's College London, London, WC2R 2LS, UK}

\date{\today}

\begin{abstract}
We consider Grand Unified Theories based on $SO(10)$ which originate from  
string/$M$ theory on $G_2$ manifolds or Calabi-Yau spaces with discrete symmetries.
In this framework we are naturally led to a novel 
solution of 
the doublet-triplet splitting problem previously considered by Dvali which involves an extra vector-like Standard Model family and light, but weakly
coupled colour triplets. These additional states are predicted to be accessible at the LHC and also induce R-parity violation. Gauge coupling unification occurs
with a larger GUT coupling.
\end{abstract}

\maketitle

\section{Introduction}

The hierarchy or naturalness problem which is the question, \emph{what physics stabilises the electroweak scale of the Standard Model (SM) to be so low ($\mathcal{O}$(100 GeV))?}, has become even sharper after run I of the CERN Large Hadron Collider (LHC). A remarkably Standard Model-like Higgs boson was discovered with a mass around 125 GeV, with no evidence for new physics beyond the Standard Model \cite{Aad:2012tfa,Chatrchyan:2012ufa,CMS-PAS-HIG-14-009}. Supersymmetry in principle solves this problem, but the limits from run I of the LHC can be of order a few TeV for the superparticle masses \cite{Aad:2014lra,Chatrchyan:2013lya,Aad:2014nua}, whilst naturalness arguments suggest that such particles would have been seen by now. 

Since many of the results and limits from LHC searches have been in the context of the Minimal Supersymmetric Standard Model (MSSM) or very special, simple, subsets of the MSSM parameter space, one could ask: what limits would we obtain in more general supersymmetric models? However, since there are literally infinite choices to be made in constructing such models, we might first ask, what reasonable guides do we have to go beyond the MSSM? 

Results over the past decade or so have shown that the simple combination of supersymmetry breaking moduli stabilisation and string/$M$ theory can in fact be a very useful guide to constructing models \cite{Acharya:2007rc,Acharya:2008zi,Acharya:2012tw,Kachru:2003aw,Balasubramanian:2005zx}. Namely, the progress in understanding supersymmetry breaking and moduli stabilisation in string/$M$ theory has been shown to lead to effective models with distinctive features and very few parameters. 

One is thus led to consider supersymmetric grand unified theories (GUTs)
based on simple groups, such as $SU(5)$ 
which explain the fermion quantum numbers and unify the three Standard Model forces, in the string/$M$ theory context. 
In doing so, however, we have to face the basic problem of GUTs -- the Higgs doublet-triplet splitting problem: the Standard Model Higgs doublet is unified into a GUT multiplet containing colour triplets which can mediate proton decay too quickly. In many models, including those originating in string/$M$ theory, this problem is often solved by making the colour-triplets very massive \cite{Witten:1985xc,Mohapatra:2007vd,Lee:2010gv}, something often achieved with a discrete symmetry whose effective action on the triplets is different from that on the doublets.  

The main purpose of this Letter is to extend the scope of the $M$ theory approach 
from the previously considered $SU(5)$/MSSM case arising from $M$ theory on $G_2$ manifolds \cite{Witten:2001bf,Acharya:2008zi}
to $SO(10)$, where an entire fermion family 
$Q,u^c,d^c,L,e^c,N$, including a charge conjugated right-handed neutrino $N$, is unified within a single $\mathbf{16}$ representation denoted $\mathbf{16}^m$.
In particular we focus on the Higgs doublet-triplet splitting problem, whose solution turns out to be necessarily
quite different in the $SO(10)$ case, leading to distinct phenomenological 
constraints and predictions. 
In the remainder of this Letter, we first review some basic ideas and results from $M$ theory,
followed by their application in the 
$SU(5)$/MSSM context, before embarking on a discussion of the new $SO(10)$ case.

$M$ theory on a manifold of $G_2$ holonomy  \footnote{Constructions of compact $G_2$ holonomy manifolds are described in \cite{G2,2012arXiv1207.4470C}.} leads elegantly to four dimensional models with supersymmetry. In such models, both Yang-Mills fields and chiral fermions arise from very particular kinds of singularities in the extra dimensions \cite{Acharya:2001gy,Acharya:2004qe}.
Yang-Mills fields are localised along three-dimensional subspaces of the seven extra dimensions along which there is an orbifold singularity. Chiral fermions, which couple to these Yang-Mills fields, arise from additional localised points at which there is a conical singularity. Therefore, different GUT multiplets are
localised at different points in the extra dimensions. 
The GUT gauge group can be broken to $SU(3)\times SU(2) \times U(1)$ (possibly with additional $U(1)$ factors) by Wilson lines on the three-dimensional subspace supporting the gauge fields.
Compact manifolds of $G_2$ holonomy -- being Ricci flat and having a finite fundamental group -- can not have continuous symmetries, but could have discrete symmetries. If present, such symmetries play a very important role in the physics. In particular, for the $SU(5)$ case, Witten showed that such symmetries can solve the doublet-triplet splitting problem\cite{Witten:2001bf}. 

\section{$SU(5)$}

In Witten's $M$ theory approach to $SU(5)$, the combination of the discrete symmetry, the Wilson lines and the fact that GUT multiplets are localised at points, allows
one to prevent the MSSM Higgs doublets, $H_u$ and $H_d$, from having a mass (the $\mu$-term) whilst the colour triplets $D$ and $\overline{D}$ could have large masses. For simplicity we assume that the symmetry is ${\bf Z_N}$. We use the following notation: 
$\overline{\mathbf{5}}^w$ is the multiplet containing $H_d$ and  $\overline{D}$ and is localised along the Wilson line (which is a circle in the extra dimensions); 
${\bf 5}^h$ is the multiplet containing $H_u$; ${\bf {\overline 5}}^m$ and ${\bf 10}^m$ are the matter multiplets. Then the transformation rules for these multiplets under ${\bf Z_N}$ are:
\begin{align}
\overline{\mathbf{5}}^w &\to  \eta^{\omega} \left( \eta^{\delta} H^w_d \oplus \eta^{\gamma} \overline{D}^w \right) ,\nonumber \\
\mathbf{5}^h &\to \eta^{\chi} \mathbf{5}^h , \label{eq:w5} \\
\overline{\mathbf{5}}^m &\to \eta^{\tau} \overline{\mathbf{5}}^m , \nonumber \\
\mathbf{10}^m &\to \eta^{\sigma} \mathbf{10}^m , \nonumber
\end{align}
where $\eta \equiv e^{2\pi i /N}$, $2\delta + 3\gamma = 0$ mod $N$ .
By requiring that Yukawa couplings, Majorana neutrino masses, and colour-triplet masses must be present, we obtain constraints on the charges as can be seen in Table \ref{tab:g2mssm} where we chose $\omega=0$.
\begin{table}[h]
\caption{\label{tab:g2mssm}  Couplings and charges for $SU(5)$ operators.}
\begin{ruledtabular}
\begin{tabular}{cc}
Coupling & Constraint   \\ \hline
$H_u^h \mathbf{10}^m \mathbf{10}^m$ & $2 \sigma + \chi = 0$ mod $N$ \\ 
$H_d^w \mathbf{10}^m \overline{\mathbf{5}}^m$ & $\sigma + \tau +\delta = 0$ mod $N$ \\ 
$ H_u^w H_u^w  \overline{\mathbf{5}}^m  \overline{\mathbf{5}}^m$ & $2\chi + 2\tau = 0$ mod $N$ \\ 
$\overline{D}^w D^h$ & $\chi + \gamma = 0$ mod $N$\\ 
\end{tabular}
\end{ruledtabular}
\end{table}

One can solve these by writing all angles in terms of, say, $\sigma$
\begin{align}
\chi &= -\gamma = -2 \sigma \mod N, \nonumber \\
\delta &= -3\sigma + N/2 \mod N , \\
\tau &= 2\sigma + N/2 \mod N, \nonumber
\end{align}
which {\it automatically forbids the $\mu$-term and dimension four and five proton decay operators.}

The discrete symmetry forces $\mu=0$, however
phenomenologically, $\mu \geq$ $\mathcal{O}$(100) GeV from direct limits on the masses of charged Higgsinos from colliders. The symmetry must therefore be broken. 
Since the discrete symmetry is a geometric symmetry of the extra dimensions, the moduli fields are naturally charged under it. Moduli stabilisation for $G_2$-manifolds was considered in \cite{Acharya:2008hi}, and it was shown that (asymptotically free) gauge interactions in the hidden sector can generate a moduli potential capable of spontaneously breaking supersymmetry at a hierarchically small scale which {\it stabilises all the moduli}.  A key point behind the success of this mechanism and which plays a crucial role in the following, is that, in $M$ theory compactifications on $G_2$-manifolds {\it without} fluxes, all of the moduli fields $s_i$ reside in chiral superfields which contain axions. The shift symmetries enjoyed by these axions, combined with holomorphy, prevent terms in the superpotential which are polynomial in the moduli \cite{Acharya:2004qe}.

Generically the vacua of the potential will spontaneously break the ${\bf Z_N}$ symmetry. This then generates an effective $\mu$ term from, e.g. K\"{a}hler potential operators of the form
\beq
K \supset \frac{s}{m_{pl}} H_u H_d \;\;\;+\;\;h.c.,
\eeq
\`a la Giudice-Masiero \cite{Giudice:1988yz}, where $s$ generically denotes a modulus field of the appropriate charge and $m_{pl}$ is the Planck scale. Note that such terms are forbidden in the superpotential due to holomorphy and the axion shift symmetries.
From \cite{Acharya:2007rc,Acharya:2008zi,Acharya:2008hi,Acharya:2012tw} we know that the moduli vevs are approximately $\langle s \rangle \sim 0.1 m_{pl}$, $\langle F_s \rangle \sim m_{1/2} m_{pl} $ and from the standard supergravity Lagrangian \cite{Brignole:1997dp} we get an effective $\mu$-term:
\beq
\mu = \langle m_{3/2} K_{H_u H_d} - F^k K_{H_u H_d k} \label{eq:muterms}\rangle ,
\eeq 
which leads to
\beq
\mu \sim \frac{\langle s \rangle}{m_{pl}} m_{3/2} + \frac{\langle F_{s} \rangle }{m_{pl}} .
\eeq

Since gaugino masses are suppressed \cite{Acharya:2007rc,Acharya:2008zi,Acharya:2008hi,Acharya:2012tw}, the $F$-term vev is subleading and we get
\beq
\mu \sim 0.1 m_{3/2} \sim \mathcal{O}(TeV) \ .
\eeq

\section{$SO(10)$}

Following this recap,  
we now turn to our $M$ theory approach to $SO(10)$, where a novel solution to the doublet-triplet
splitting problem seems to be required. Since the Wilson line is in the adjoint representation, it can break $SO(10)$ to $SU(3)\times SU(2)\times U(1)_Y \times U(1)$ and the Wilson line itself
is a combination of $U(1)_Y$ and the additional $U(1)$. If we consider a fundamental of $SO(10)$ localised along a Wilson line, then its
transformation properties under the ${\bf Z_N}$ symmetry are in analogy to 
Eq. (\ref{eq:w5}),
\begin{equation}\label{eq:W10}
\mathbf{10}^w \to \eta^{\omega} \left(\eta^{-\alpha} H_d^w  \oplus \eta^{\beta} \overline{D}^w 
\oplus  \eta^{\alpha} H_u^w \oplus  \eta^{-\beta} D^w
\right).
\end{equation}
In minimal $SO(10)$ the $\mu$-term arises from a term in the superpotential of the form 
\beq
W \supset \mu \mathbf{10}^w\mathbf{10}^w = \mu \left(H^w_u H^w_d + D^w  \overline{D}^w\right) ,
\eeq

The discrete symmetry will forbid this term in a general model. As in the $SU(5)$ case,
once the symmetry is broken by moduli vevs, the term will be generated in the K\"{a}hler potential
via the Giudice-Masiero mechanism. This will give a $\mu$-term at the TeV scale, which, in $SO(10)$
also generates a similar mass for the triplet $D$.


In Appendix A  we consider the effect of adding additional ${\bf 10}$ multiplets and show that 
this does not help. Henceforth we shall only consider a single light
$\mathbf{10}^w $, without any extra ${\bf 10}$ multiplets at low energies.

Assuming a single light $\mathbf{10}^w $, it is possible to use the discrete symmetry to forbid certain couplings, 
namely to {\it decouple $D^w$ and $\overline{D}^w$ from matter}.
Such couplings arise from the operator $\mathbf{10}^w \mathbf{16}^m \mathbf{16}^m$, with 
$\mathbf{16}^m$ denoting the three $SO(10)$ multiplets, each containing a SM family plus right handed neutrino $N$.
If $\mathbf{16}^m$ transforms as $\eta^{\kappa} \mathbf{16}^m$, the couplings and charge constraints are in Table \ref{tab:SO10}, where we allow for up-type quark Yukawa couplings together with couplings to the right-handed neutrinos,
\beq
y_u^{ij}H_u^w \mathbf{16}^m_i \mathbf{16}^m_j \equiv y_u^{ij}H_u^w (Q_iu_j^c+L_iN_j +i\leftrightarrow j),
\label{yu}
\eeq
and similarly for down-type quarks and charged leptons.

Explicit examples realising these conditions will be given later.

\begin{table}[h]
\caption{\label{tab:SO10}  Couplings and charges for $SO(10)$ operators.}
\begin{ruledtabular}
\begin{tabular}{cc}
Coupling & Constraint \\ \hline
$H_u^w \mathbf{16}^m \mathbf{16}^m$ & $2\kappa + \alpha + \omega = 0$ mod $N$ \\ 
$H_d^w \mathbf{16}^m \mathbf{16}^m$ & $2\kappa - \alpha + \omega = 0$ mod $N$ \\ 
$D^w \mathbf{16}^m \mathbf{16}^m$ & $2\kappa - \beta + \omega \neq 0$ mod $N$ \\
$\overline{D}^w \mathbf{16}^m \mathbf{16}^m$ & $2\kappa + \beta + \omega \neq 0$ mod $N$\\ 
\end{tabular}
\end{ruledtabular}
\end{table}

The suppression of colour triplet couplings to matter 
was previously considered by Dvali in \cite{Dvali:1995hp} and also \cite{Kilian:2006hh,Reuter:2007eh,Howl:2007zi} from a bottom-up perspective.  


Next we consider the breaking of the discrete symmetry via the moduli vevs as discussed above,
leading to proton decay. 
For proton decay, the relevant operators can be generated in the K\"{a}hler potential, schematically,
writing $D=D^w$,
\beqa
K &\supset \frac{s}{m_{pl}^2} D Q Q +\frac{s}{m_{pl}^2} D e^c u^c + \frac{s}{m_{pl}^2} D N d^c + \nonumber \\
& + \frac{s}{m_{pl}^2}\overline{D} d^c u^c + \frac{s}{m_{pl}^2}\overline{D} Q L .
\eeqa
Just like the $\mu$-term, the effective superpotential may be calculated from supergravity to be
\beqa
W_{eff} & \supset \lambda D Q Q  + \lambda D e^c u^c + \lambda D N d^c\label{eq:proton} + \nonumber \\
& +\lambda \overline{D} d^c u^c+\lambda \overline{D} Q L, 
\eeqa
where
\beq
\lambda \approx \frac{1}{m^2_{pl}}\left( \langle s \rangle m_{3/2} + \langle F_{s}\rangle  \right) \sim 10^{-14} . \label{eq:lambda}
\eeq

Notice that unlike the case of SU(5), there is no SO(10) invariant bilinear term $\kappa L H_u$ whose presence would lead to fast proton decay. We estimate the scalar triplet induced proton decay rate to be 
\beq
\Gamma_p \approx \frac{\left| \lambda^2 \right|^2}{16 \pi^2}\frac{m_p^5}{m_D^4} .
\eeq
Generically, the mass of the colour triplets is of the same order as $\mu$, i.e., $m_D \sim 10^3$ GeV, so 
the proton lifetime is
\beq
\tau_p= \Gamma_p^{-1} \sim 10^{38} \;\mbox{yrs} ,
\eeq
which exceeds the current experimental limit.

Now consider the $D$ triplet decay rate:
\beq
\Gamma_D \sim \lambda^2 m_D \sim ( 0.1 \;\mbox{sec})^{-1} .
\eeq
The associated lifetime of 0.1 sec is (just) short enough to be consistent with BBN constraint. 
They will also give interesting collider signatures due to their long-lived nature.

\subsection{Vector-like Family}

Gauge coupling unification is in general spoiled by light colour triplets, unless they 
are also accompanied by additional light doublet states. 
In the present framework,
the only way we know of to circumvent this issue is the presence of light additional states which complete the triplets into complete GUT multiplets.
Happily, this can also be achieved by use of the discrete symmetry.
First we introduce a 
vector-like pair of ${\bf 16}$'s, labelled as ${\bf 16}_X+{\bf {\overline{16}}}_X$. Next
a GUT-scale mass is given to their colour triplet components $d^c_X,\overline{d^c}_X$
whilst keeping the remaining particles light. Suitable charges under the discrete symmetry can forbid the appropriate mass terms and the large mass
can arise from membrane instantons if the ${\bf 16}_X$ and ${\bf {\overline{16}}}_X$ are close by on the $G_2$ manifold \cite{Witten:1985xc}.

We take ${\bf 16}_X$ to be localised along a Wilson line, and find that it
transforms under the discrete symmetry as
\begin{align}
{\bf 16}_X \to & \eta^x \left( \eta^{-3\gamma} L \oplus \eta^{ 3\gamma+\delta} e^c \oplus  \eta^{3 \gamma - \delta} N \oplus  \eta^{-\gamma-\delta} u^c \oplus \right. \nonumber \\
& \left.  \oplus \  \eta^{-\gamma +\delta} d^c \oplus  \eta^{\gamma} Q \right) .
\end{align}
Assuming ${\bf {\overline{16}}}_X$ transforms without Wilson line phases, ${\bf {\overline{16}}}_X \to \eta^{\overline x}\,  {\bf {\overline{16}}}_X$, the condition for the mass term is 
\begin{equation}
  \overline{d^c}_X d^c_X : x - \gamma + \delta + \overline{x} = 0 \mod N ,\
  \label{eq:MassivedX}
\end{equation}
whilst forbidding all the other self couplings that would arise from ${\bf 16}_X {\bf {\overline{16}}}_X$.

The light $D^w$ and $\overline{D}^w$ from the original ${\bf 10}^w$ then ``complete" the 
${\bf 16}_X+{\bf {\overline{16}}}_X$ pair,
since they have the same SM quantum numbers as the missing $d^c_X,\overline{d^c}_X$. The light states in the ${\bf 16}_X$ and ${\bf \overline{16}_X}$ also
obtain masses via the K\"{a}hler potential of order a TeV via the Giudice-Masiero mechanism.
Gauge unification is clearly restored, albeit with a larger gauge coupling
at the GUT scale due to the extra low energy matter content (relative to the MSSM).

Effective $\mu$-terms induced by moduli vevs of the form $\mu {\bf 16}^m \overline{\bf 16}_X$ are then generated and one might be concerned about too much mixing
with quarks and leptons. However, one finds that all the light components of the extra matter decouple from ordinary matter, with mixings supressed by terms
of order (\ref{eq:lambda}).
For example, consider the up-type quark sector. The superpotential contribution to the mass matrix is, schematically,
$U_L \cdot M_u \cdot U_R$
, with
$U_L = \begin{pmatrix}
u^i & u_X & \overline{u^c}_X
\end{pmatrix}^T$,
$U_R = \begin{pmatrix}
(u^c)^i & \overline{u}_X & u^c_X 
\end{pmatrix}^T$
, and 

\begin{equation}
M_u=
\begin{pmatrix}
y_u^{ij} \langle H_u \rangle & \mu^i_{XQ}& 0  \\
0 & \mu_{XXQ}& 0   \\
\mu^j_{Xu} & 0 & \mu_{Xu} 
\end{pmatrix}
\label{eq:massMatrix}
\end{equation}



Here $\mu^i_{XQ},\mu^j_{Xu},\mu_{XXQ}$, $\mu_{Xu}$ are moduli induced $\mu$-type parameters of $\mathcal{O}(\mbox{TeV})$ while the vanishing entries are non-zero only to first order in moduli-induced trilinear interactions that are vanishingly small, $\mathcal{O}(10^{-14})$. We have found numerically that flavour changing neutral currents (FCNCs) are highly suppressed by this structure. This can be understood analytically
in the approximation that the electroweak masses can be ignored,
since $y_u \langle H_u\rangle /\mu \sim \mathcal{O}(0.1)$.
In this approximation, the third lightest $u$-quark will be given by the two component Weyl quarks
\begin{align}
t=u_3^\prime & \simeq \frac{1}{\sqrt{(\mu_X^3)^2+(\mu_{XXQ})^2}} ((\mu_{XXQ}) u_3 - (\mu_X^3) u_X) ,\\
t^c=(u^c_3)^\prime & \simeq \frac{1}{\sqrt{(\mu^3_{Xu})^2+(\mu_{Xu})^2}} ((\mu_{Xu} (u^c)_3 - (\mu^3_{Xu})  u^c_X) ,
\end{align}
and as a result the light $up$-quark, which we denote $t$, does not result in a mixing including $\overline{u^c}_X$. This is important, since  $\overline{u^c}_X$  in $U_L$  couples to $Z$ differently, only through the electromagnetic contribution to the neutral current and not via the $J^\mu_3$ contribution. Consequently, 
FCNCs will be naturally suppressed 
and the CKM matrix should have only small deviations from unitarity.
Furthermore we note that the resulting matter states couple to the Higgses and $Z$ as in the MSSM.


\subsection{See-saw Mechanism}

Introducing the $\mathbf{16}_X$ and $\overline{\bf 16}_{X}$ will play a crucial role in breaking the extra $U(1)$ subgroup of $SO(10)$ and generating right-handed neutrino masses. 
We assume that a mechanism similar to the one proposed by Kolda-Martin \cite{Kolda:1995iw} is in effect, such that the right-handed neutrino components acquire a non-trivial high-scale vev along the D-flat direction, $\langle N_X \rangle = \langle  \overline{N}_X \rangle =v_X$, which in turn breaks the rank.
However the scale $v_X$ is constrained, as discussed below.

Presence of the $\mathbf{16}_X$ and $\overline{\bf 16}_{X}$  with vevs in their right-handed neutrino components gives us the possibility of having a see-saw mechanism for light physical neutrino masses. Such a mechanism is welcome since representations larger than the $\mathbf{45}$ are absent in $M$ theory \cite{Acharya:2001gy}.
In the present framework, a Majorana mass term for the right handed neutrino in ${\bf{16}}^m$ is generated by letting the discrete symmetry to allow the Planck suppressed operator $\frac{1}{m_{pl}}\overline{\bf 16}_{X}\overline{\bf 16}_{X}\mathbf{16}^m\mathbf{16}^m$. This requires charges to satisfy
$2 \overline{x}+2\kappa=0\mod N$, and leads to the Majorana mass $M \sim \frac{ {v_X}^2}{m_{pl}}$.


Due to the nature of $SO(10)$, the neutrinos will have the same Yukawa coupling as the up-type quarks $y_u^{ij}$, as in Eq.~(\ref{yu}),
leading to their Dirac masses being the same as the up-quark masses. For the case of the top quark mass we would need $M \sim 10^{14}$ GeV in order to give a realistic neutrino mass.
Such a high value can only be achieved by the above see-saw mechanism if $v_X \gtrsim 10^{16}$ GeV.

The magnitude of $v_X$ is also constrained by R-parity violating (RPV) dynamically generated operators, due to moduli and $N_X, \overline{N}_X$ vevs, arising from the K\"{a}hler interactions
\beq
K_{RPV} \supset  \frac{s}{m^3_{pl}} \mathbf{16}_X \mathbf{16}^m \mathbf{16}^m \mathbf{16}^m +  \frac{s}{m^2_{pl}} \mathbf{10}^w \mathbf{16}_X \mathbf{16}^m .
\eeq
Since $s$ and $N_X$ acquire vevs, these operators generate the effective superpotential terms (otherwise forbidden by the discrete symmetry),
\beqa
W^{eff}_{RPV} &\supset& \lambda \frac{v_X}{m_{pl}} L L e^c + \lambda\frac{v_X}{m_{pl}} LQ d^c +  \lambda \frac{v_X}{m_{pl}} u^c d^c d^c + \nonumber \\
& & + \lambda v_X L H_u ,  \label{eq:RPV}
\eeqa
with $\lambda \sim \mathcal{O}(10^{-14})$. One can absorb the last term into $\mu H_dH_u$
by a small rotation $\mathcal{O}(v_X/m_{pl})$
in $(H_d,L)$ space,
\beq
W^{eff}_{RPV} \supset  y_e  \frac{v_X}{m_{pl}} L L e^c +  y_d   \frac{v_X}{m_{pl}} LQ d^c +  \lambda  \frac{v_X}{m_{pl}} u^c d^c d^c ,  \label{eq:RPVrot}
\eeq
where the first two terms originate from the Yukawa couplings $y_eH_dLe^c$, etc.,
and we have dropped the $\mathcal{O}(\lambda)$ contributions to these terms since now the 
Yukawa rotated contributions are much larger.

We emphasise that there exist explicit solutions to the constraints on all of the charges and couplings that we have discussed. These are Table \ref{tab:SO10}, Eq.~(\ref{eq:MassivedX}), the Majorana mass term, suppressing the RPV operators and cross-terms between the visible matter, $16_{X}$ and $16_{\overline{X}}$ necessary for Eq.~(\ref{eq:massMatrix}). An example is given by
\beq
(N,\omega, \alpha,  \beta, \kappa, x, \gamma, \delta, \overline{x} ) = (16, 4, 0, 1, 6, 2, 1, 13, 2).
\eeq
which is also anomaly free, as can be checked by explicit calculations \cite{Araki:2008ek}.

The last term in Eq. (\ref{eq:RPV}) is the bilinear RPV operator which is contained as it contributes to neutrino masses \cite{Acharya:2014vha,Banks:1995by}. The constraint is
$ \lambda v_X \lesssim \mathcal{O}(1 \mbox{ GeV})$, which leads to the upper bound $v_X \lesssim 10^{14}$ GeV in contradiction with the see-saw requirement $v_X \sim 10^{16}$ GeV assumed in the above estimates. However, there is a natural way within this framework to further suppress the bilinear RPV terms. This happens when the charges of the moduli fields under the discrete symmetry are such that the leading order terms in $K$, which are  linear in the moduli i.e. $\frac{s}{m_{pl}^2}v_XLH_u$, are {\it forbidden} by the symmetry, with the leading term arising at higher order in the moduli. If the leading term arises at cubic order or higher, (e.g. $K \sim \frac{s^3}{m_{pl}^4} v_XLH_u$ then the suppression will be sufficient. Furthermore, some moduli may have smaller vevs than others in a detailed model, leading to additional suppression. 

The RPV terms in Eq. (\ref{eq:RPVrot}) induce the lightest supersymetric particle (LSP) decay. We can estimate its lifetime as \cite{Acharya:2011te}:
\begin{equation}
\tau_{LSP} \simeq \frac{10^{-13} \sec}{ \left( v_X/m_{pl}\right)^2}\left( \frac{m_0}{10 \ \mbox{TeV}}\right)^4 \left( \frac{100 \ \mbox{GeV}}{m_{LSP}}\right)^5 .
\end{equation}
Since, as discussed above, $v_X/m_{pl}\sim 10^{-2}$, 
one finds $\tau_{LSP} \sim 10^{-9} \sec$. This value is compatible with current bounds $\tau_{LSP} \lesssim 1$ sec \cite{Dreiner:1997uz}, from BBN. Clearly the LSP is not a good DM candidate. However, $M$ theory usually provides axion dark matter candidates \cite{Svrcek:2006yi,Acharya:2010zx}





\section{Conclusion}We have discussed the origin of an $SO(10)$ SUSY GUT from $M$ theory on a $G_2$ manifold. We were naturally led to a novel 
solution of the doublet-triplet splitting problem involving an extra ${\bf 16}_X+{\bf {\overline{16}}}_X$ vector-like pair where discrete symmetries of the extra dimensions were used to prevent proton decay by suppressing the Yukawa couplings of colour triplets.
Such models maintain gauge coupling unification but with a larger GUT coupling than 
predicted by the MSSM.
We argue that these extra multiplets, also required to break the additional $U(1)$ gauge symmetry,
inevitably lead to R-parity violating operators . 
Even though the moduli potential generically breaks the discrete symmetry, we have seen that one naturally satisfies the 
constraints from the proton lifetime and decays
affecting BBN. We also have found a consistent scenario for neutrino masses arising from the high scale 
see-saw mechanism, with sufficiently suppressed RPV contributions.
We emphasise the main prediction of this approach, namely 
light states with the quantum numbers of a ${\bf 16}_X+{\bf {\overline{16}}}_X$ vector-like pair
which might be accessible in future LHC searches.

\begin{acknowledgments}
The work of BSA is supported by UK STFC via the research grant ST/J002798/1.
Two of us (MCR and SFK) thank King's College London for hospitality.
SFK acknowledges support from the
European Union FP7 ITN-INVISIBLES (Marie Curie Actions, PITN- GA-2011-
289442). MCR acknowledges support from the FCT under the grant SFRH/BD/84234/2012.
CP is supported by the KCL NMS graduate school and ICTP Trieste.
The work of KB is supported by a KCL GTA studentship.
\end{acknowledgments}

\appendix
\section{} \label{App:AppendixA}

In this Appendix we consider the effect of adding additional ${\bf 10}$ multiplets and show that 
this does not change the argument presented in the main text for one ${\bf 10}$ multiplet.
With additional ${\bf 10}$ multiplets, one can forbid some couplings between the different members of the various ${\bf 10}$ multiplets, but one can see that there will typically be more than one pair of light Higgs doublets which tend to destroy gauge coupling unification. Consider one additional ${\bf 10}$, denoted $\mathbf{10}^h$ without Wilson line phases: $\mathbf{10}^h \rightarrow \eta^\xi \mathbf{10}^h$. We have eight possible gauge invariant couplings with a $\mathbf{10}^w$ and $\mathbf{10}^h$ that can be written in matrix form as
\begin{align}
W \supset \mathbf{H}_d^T \cdot \mu_H \cdot \mathbf{H}_u + \overline{\mathbf{D}}^T \cdot M_D \cdot \mathbf{D} ,
\end{align}
where $\mu_H$ and $M_D$ are two $2\times 2$ superpotential mass parameters matrices, $\mathbf{H}_{u,d}^T = \left(H^w_{u,d}, H^h_{u,d}\right)$, $\mathbf{\overline{D}}^T = \left (\overline{D}^w,\overline{D}^h\right)$, and $\mathbf{D}^T = \left(D^w,D^h\right)$. The entries of the matrices are non-vanishing depending on which of the following discrete charge combinations are zero (mod $N$)
\begin{align}
D^w\overline D^w , \ H^w_u H^w_d & : \  2 \omega   , \nonumber \\
D^h\overline D^h , \  H^h_u  H^h_d & : \ 2 \xi   ,\nonumber \\
H^w_u  H^h_d &: \ \alpha + \omega + \xi   , \label{eq:10w10m} \\
H^h_u  H^w_d &: \ -\alpha +\omega+\xi   , \nonumber\\
D^w \overline D^h &: \ -\beta +\omega + \xi    , \nonumber\\
D^h \overline D^w &: \ \beta +\omega +\xi  .\nonumber
\end{align}
The naive doublet-triplet splitting solution would be for $\mu_H$ to have only one zero eigenvalue, with $M_D$ having all non-zero eigenvalues. One finds that there is no choice of constraints in Eq. (\ref{eq:10w10m}) that accomplishes this.

\bibliographystyle{apsrev4-1}
\bibliography{so10g2_arxiv}

\begin{thebibliography}{35}%
\makeatletter
\providecommand \@ifxundefined [1]{%
 \@ifx{#1\undefined}
}%
\providecommand \@ifnum [1]{%
 \ifnum #1\expandafter \@firstoftwo
 \else \expandafter \@secondoftwo
 \fi
}%
\providecommand \@ifx [1]{%
 \ifx #1\expandafter \@firstoftwo
 \else \expandafter \@secondoftwo
 \fi
}%
\providecommand \natexlab [1]{#1}%
\providecommand \enquote  [1]{``#1''}%
\providecommand \bibnamefont  [1]{#1}%
\providecommand \bibfnamefont [1]{#1}%
\providecommand \citenamefont [1]{#1}%
\providecommand \href@noop [0]{\@secondoftwo}%
\providecommand \href [0]{\begingroup \@sanitize@url \@href}%
\providecommand \@href[1]{\@@startlink{#1}\@@href}%
\providecommand \@@href[1]{\endgroup#1\@@endlink}%
\providecommand \@sanitize@url [0]{\catcode `\\12\catcode `\$12\catcode
  `\&12\catcode `\#12\catcode `\^12\catcode `\_12\catcode `\%12\relax}%
\providecommand \@@startlink[1]{}%
\providecommand \@@endlink[0]{}%
\providecommand \url  [0]{\begingroup\@sanitize@url \@url }%
\providecommand \@url [1]{\endgroup\@href {#1}{\urlprefix }}%
\providecommand \urlprefix  [0]{URL }%
\providecommand \Eprint [0]{\href }%
\providecommand \doibase [0]{http://dx.doi.org/}%
\providecommand \selectlanguage [0]{\@gobble}%
\providecommand \bibinfo  [0]{\@secondoftwo}%
\providecommand \bibfield  [0]{\@secondoftwo}%
\providecommand \translation [1]{[#1]}%
\providecommand \BibitemOpen [0]{}%
\providecommand \bibitemStop [0]{}%
\providecommand \bibitemNoStop [0]{.\EOS\space}%
\providecommand \EOS [0]{\spacefactor3000\relax}%
\providecommand \BibitemShut  [1]{\csname bibitem#1\endcsname}%
\let\auto@bib@innerbib\@empty
\bibitem [{\citenamefont {Aad}\ \emph {et~al.}(2012)\citenamefont {Aad} \emph
  {et~al.}}]{Aad:2012tfa}%
  \BibitemOpen
  \bibfield  {author} {\bibinfo {author} {\bibfnamefont {G.}~\bibnamefont
  {Aad}} \emph {et~al.} (\bibinfo {collaboration} {ATLAS Collaboration}),\
  }\href {\doibase 10.1016/j.physletb.2012.08.020} {\bibfield  {journal}
  {\bibinfo  {journal} {Phys.Lett.}\ }\textbf {\bibinfo {volume} {B716}},\
  \bibinfo {pages} {1} (\bibinfo {year} {2012})},\ \Eprint
  {http://arxiv.org/abs/1207.7214} {arXiv:1207.7214 [hep-ex]} \BibitemShut
  {NoStop}%
\bibitem [{\citenamefont {Chatrchyan}\ \emph {et~al.}(2012)\citenamefont
  {Chatrchyan} \emph {et~al.}}]{Chatrchyan:2012ufa}%
  \BibitemOpen
  \bibfield  {author} {\bibinfo {author} {\bibfnamefont {S.}~\bibnamefont
  {Chatrchyan}} \emph {et~al.} (\bibinfo {collaboration} {CMS Collaboration}),\
  }\href {\doibase 10.1016/j.physletb.2012.08.021} {\bibfield  {journal}
  {\bibinfo  {journal} {Phys.Lett.}\ }\textbf {\bibinfo {volume} {B716}},\
  \bibinfo {pages} {30} (\bibinfo {year} {2012})},\ \Eprint
  {http://arxiv.org/abs/1207.7235} {arXiv:1207.7235 [hep-ex]} \BibitemShut
  {NoStop}%
\bibitem [{CMS(2014)}]{CMS-PAS-HIG-14-009}%
  \BibitemOpen
  \href@noop {} {\emph {\bibinfo {title} {{Precise determination of the mass of
  the Higgs boson and studies of the compatibility of its couplings with the
  standard model}}}},\ \bibinfo {type} {Tech. Rep.}\ \bibinfo {number}
  {CMS-PAS-HIG-14-009}\ (\bibinfo  {institution} {CERN},\ \bibinfo {address}
  {Geneva},\ \bibinfo {year} {2014})\BibitemShut {NoStop}%
\bibitem [{\citenamefont {Aad}\ \emph {et~al.}(2014{\natexlab{a}})\citenamefont
  {Aad} \emph {et~al.}}]{Aad:2014lra}%
  \BibitemOpen
  \bibfield  {author} {\bibinfo {author} {\bibfnamefont {G.}~\bibnamefont
  {Aad}} \emph {et~al.} (\bibinfo {collaboration} {ATLAS Collaboration}),\
  }\href {\doibase 10.1007/JHEP10(2014)024} {\bibfield  {journal} {\bibinfo
  {journal} {JHEP}\ }\textbf {\bibinfo {volume} {1410}},\ \bibinfo {pages} {24}
  (\bibinfo {year} {2014}{\natexlab{a}})},\ \Eprint
  {http://arxiv.org/abs/1407.0600} {arXiv:1407.0600 [hep-ex]} \BibitemShut
  {NoStop}%
\bibitem [{\citenamefont {Chatrchyan}\ \emph {et~al.}(2013)\citenamefont
  {Chatrchyan} \emph {et~al.}}]{Chatrchyan:2013lya}%
  \BibitemOpen
  \bibfield  {author} {\bibinfo {author} {\bibfnamefont {S.}~\bibnamefont
  {Chatrchyan}} \emph {et~al.} (\bibinfo {collaboration} {CMS Collaboration}),\
  }\href {\doibase 10.1140/epjc/s10052-013-2568-6} {\bibfield  {journal}
  {\bibinfo  {journal} {Eur.Phys.J.}\ }\textbf {\bibinfo {volume} {C73}},\
  \bibinfo {pages} {2568} (\bibinfo {year} {2013})},\ \Eprint
  {http://arxiv.org/abs/1303.2985} {arXiv:1303.2985 [hep-ex]} \BibitemShut
  {NoStop}%
\bibitem [{\citenamefont {Aad}\ \emph {et~al.}(2014{\natexlab{b}})\citenamefont
  {Aad} \emph {et~al.}}]{Aad:2014nua}%
  \BibitemOpen
  \bibfield  {author} {\bibinfo {author} {\bibfnamefont {G.}~\bibnamefont
  {Aad}} \emph {et~al.} (\bibinfo {collaboration} {ATLAS Collaboration}),\
  }\href {\doibase 10.1007/JHEP04(2014)169} {\bibfield  {journal} {\bibinfo
  {journal} {JHEP}\ }\textbf {\bibinfo {volume} {1404}},\ \bibinfo {pages}
  {169} (\bibinfo {year} {2014}{\natexlab{b}})},\ \Eprint
  {http://arxiv.org/abs/1402.7029} {arXiv:1402.7029 [hep-ex]} \BibitemShut
  {NoStop}%
\bibitem [{\citenamefont {Acharya}\ \emph {et~al.}(2007)\citenamefont
  {Acharya}, \citenamefont {Bobkov}, \citenamefont {Kane}, \citenamefont
  {Kumar},\ and\ \citenamefont {Shao}}]{Acharya:2007rc}%
  \BibitemOpen
  \bibfield  {author} {\bibinfo {author} {\bibfnamefont {B.~S.}\ \bibnamefont
  {Acharya}}, \bibinfo {author} {\bibfnamefont {K.}~\bibnamefont {Bobkov}},
  \bibinfo {author} {\bibfnamefont {G.~L.}\ \bibnamefont {Kane}}, \bibinfo
  {author} {\bibfnamefont {P.}~\bibnamefont {Kumar}}, \ and\ \bibinfo {author}
  {\bibfnamefont {J.}~\bibnamefont {Shao}},\ }\href {\doibase
  10.1103/PhysRevD.76.126010} {\bibfield  {journal} {\bibinfo  {journal}
  {Phys.Rev.}\ }\textbf {\bibinfo {volume} {D76}},\ \bibinfo {pages} {126010}
  (\bibinfo {year} {2007})},\ \Eprint {http://arxiv.org/abs/hep-th/0701034}
  {arXiv:hep-th/0701034 [hep-th]} \BibitemShut {NoStop}%
\bibitem [{\citenamefont {Acharya}\ \emph {et~al.}(2008)\citenamefont
  {Acharya}, \citenamefont {Bobkov}, \citenamefont {Kane}, \citenamefont
  {Shao},\ and\ \citenamefont {Kumar}}]{Acharya:2008zi}%
  \BibitemOpen
  \bibfield  {author} {\bibinfo {author} {\bibfnamefont {B.~S.}\ \bibnamefont
  {Acharya}}, \bibinfo {author} {\bibfnamefont {K.}~\bibnamefont {Bobkov}},
  \bibinfo {author} {\bibfnamefont {G.~L.}\ \bibnamefont {Kane}}, \bibinfo
  {author} {\bibfnamefont {J.}~\bibnamefont {Shao}}, \ and\ \bibinfo {author}
  {\bibfnamefont {P.}~\bibnamefont {Kumar}},\ }\href {\doibase
  10.1103/PhysRevD.78.065038} {\bibfield  {journal} {\bibinfo  {journal}
  {Phys.Rev.}\ }\textbf {\bibinfo {volume} {D78}},\ \bibinfo {pages} {065038}
  (\bibinfo {year} {2008})},\ \Eprint {http://arxiv.org/abs/0801.0478}
  {arXiv:0801.0478 [hep-ph]} \BibitemShut {NoStop}%
\bibitem [{\citenamefont {Acharya}\ \emph {et~al.}(2012)\citenamefont
  {Acharya}, \citenamefont {Kane},\ and\ \citenamefont
  {Kumar}}]{Acharya:2012tw}%
  \BibitemOpen
  \bibfield  {author} {\bibinfo {author} {\bibfnamefont {B.~S.}\ \bibnamefont
  {Acharya}}, \bibinfo {author} {\bibfnamefont {G.}~\bibnamefont {Kane}}, \
  and\ \bibinfo {author} {\bibfnamefont {P.}~\bibnamefont {Kumar}},\ }\href
  {\doibase 10.1142/S0217751X12300128} {\bibfield  {journal} {\bibinfo
  {journal} {Int.J.Mod.Phys.}\ }\textbf {\bibinfo {volume} {A27}},\ \bibinfo
  {pages} {1230012} (\bibinfo {year} {2012})},\ \Eprint
  {http://arxiv.org/abs/1204.2795} {arXiv:1204.2795 [hep-ph]} \BibitemShut
  {NoStop}%
\bibitem [{\citenamefont {Kachru}\ \emph {et~al.}(2003)\citenamefont {Kachru},
  \citenamefont {Kallosh}, \citenamefont {Linde},\ and\ \citenamefont
  {Trivedi}}]{Kachru:2003aw}%
  \BibitemOpen
  \bibfield  {author} {\bibinfo {author} {\bibfnamefont {S.}~\bibnamefont
  {Kachru}}, \bibinfo {author} {\bibfnamefont {R.}~\bibnamefont {Kallosh}},
  \bibinfo {author} {\bibfnamefont {A.~D.}\ \bibnamefont {Linde}}, \ and\
  \bibinfo {author} {\bibfnamefont {S.~P.}\ \bibnamefont {Trivedi}},\ }\href
  {\doibase 10.1103/PhysRevD.68.046005} {\bibfield  {journal} {\bibinfo
  {journal} {Phys.Rev.}\ }\textbf {\bibinfo {volume} {D68}},\ \bibinfo {pages}
  {046005} (\bibinfo {year} {2003})},\ \Eprint
  {http://arxiv.org/abs/hep-th/0301240} {arXiv:hep-th/0301240 [hep-th]}
  \BibitemShut {NoStop}%
\bibitem [{\citenamefont {Balasubramanian}\ \emph {et~al.}(2005)\citenamefont
  {Balasubramanian}, \citenamefont {Berglund}, \citenamefont {Conlon},\ and\
  \citenamefont {Quevedo}}]{Balasubramanian:2005zx}%
  \BibitemOpen
  \bibfield  {author} {\bibinfo {author} {\bibfnamefont {V.}~\bibnamefont
  {Balasubramanian}}, \bibinfo {author} {\bibfnamefont {P.}~\bibnamefont
  {Berglund}}, \bibinfo {author} {\bibfnamefont {J.~P.}\ \bibnamefont
  {Conlon}}, \ and\ \bibinfo {author} {\bibfnamefont {F.}~\bibnamefont
  {Quevedo}},\ }\href {\doibase 10.1088/1126-6708/2005/03/007} {\bibfield
  {journal} {\bibinfo  {journal} {JHEP}\ }\textbf {\bibinfo {volume} {0503}},\
  \bibinfo {pages} {007} (\bibinfo {year} {2005})},\ \Eprint
  {http://arxiv.org/abs/hep-th/0502058} {arXiv:hep-th/0502058 [hep-th]}
  \BibitemShut {NoStop}%
\bibitem [{\citenamefont {Witten}(1985)}]{Witten:1985xc}%
  \BibitemOpen
  \bibfield  {author} {\bibinfo {author} {\bibfnamefont {E.}~\bibnamefont
  {Witten}},\ }\href {\doibase 10.1016/0550-3213(85)90603-0} {\bibfield
  {journal} {\bibinfo  {journal} {Nucl.Phys.}\ }\textbf {\bibinfo {volume}
  {B258}},\ \bibinfo {pages} {75} (\bibinfo {year} {1985})}\BibitemShut
  {NoStop}%
\bibitem [{\citenamefont {Mohapatra}\ and\ \citenamefont
  {Ratz}(2007)}]{Mohapatra:2007vd}%
  \BibitemOpen
  \bibfield  {author} {\bibinfo {author} {\bibfnamefont {R.~N.}\ \bibnamefont
  {Mohapatra}}\ and\ \bibinfo {author} {\bibfnamefont {M.}~\bibnamefont
  {Ratz}},\ }\href {\doibase 10.1103/PhysRevD.76.095003} {\bibfield  {journal}
  {\bibinfo  {journal} {Phys.Rev.}\ }\textbf {\bibinfo {volume} {D76}},\
  \bibinfo {pages} {095003} (\bibinfo {year} {2007})},\ \Eprint
  {http://arxiv.org/abs/0707.4070} {arXiv:0707.4070 [hep-ph]} \BibitemShut
  {NoStop}%
\bibitem [{\citenamefont {Lee}\ \emph {et~al.}(2011)\citenamefont {Lee},
  \citenamefont {Raby}, \citenamefont {Ratz}, \citenamefont {Ross},
  \citenamefont {Schieren} \emph {et~al.}}]{Lee:2010gv}%
  \BibitemOpen
  \bibfield  {author} {\bibinfo {author} {\bibfnamefont {H.~M.}\ \bibnamefont
  {Lee}}, \bibinfo {author} {\bibfnamefont {S.}~\bibnamefont {Raby}}, \bibinfo
  {author} {\bibfnamefont {M.}~\bibnamefont {Ratz}}, \bibinfo {author}
  {\bibfnamefont {G.~G.}\ \bibnamefont {Ross}}, \bibinfo {author}
  {\bibfnamefont {R.}~\bibnamefont {Schieren}},  \emph {et~al.},\ }\href
  {\doibase 10.1016/j.physletb.2010.10.038} {\bibfield  {journal} {\bibinfo
  {journal} {Phys.Lett.}\ }\textbf {\bibinfo {volume} {B694}},\ \bibinfo
  {pages} {491} (\bibinfo {year} {2011})},\ \Eprint
  {http://arxiv.org/abs/1009.0905} {arXiv:1009.0905 [hep-ph]} \BibitemShut
  {NoStop}%
\bibitem [{\citenamefont {Witten}()}]{Witten:2001bf}%
  \BibitemOpen
  \bibfield  {author} {\bibinfo {author} {\bibfnamefont {E.}~\bibnamefont
  {Witten}},\ }\href@noop {} {\ }\Eprint {http://arxiv.org/abs/hep-ph/0201018}
  {arXiv:hep-ph/0201018 [hep-ph]} \BibitemShut {NoStop}%
\bibitem [{Note1()}]{Note1}%
  \BibitemOpen
  \bibinfo {note} {Constructions of compact $G_2$ holonomy manifolds are
  described in \cite {G2,2012arXiv1207.4470C}.}\BibitemShut {Stop}%
\bibitem [{\citenamefont {Acharya}\ and\ \citenamefont
  {Witten}(2001)}]{Acharya:2001gy}%
  \BibitemOpen
  \bibfield  {author} {\bibinfo {author} {\bibfnamefont {B.~S.}\ \bibnamefont
  {Acharya}}\ and\ \bibinfo {author} {\bibfnamefont {E.}~\bibnamefont
  {Witten}},\ }\href@noop {} {\  (\bibinfo {year} {2001})},\ \Eprint
  {http://arxiv.org/abs/hep-th/0109152} {arXiv:hep-th/0109152 [hep-th]}
  \BibitemShut {NoStop}%
\bibitem [{\citenamefont {Acharya}\ and\ \citenamefont
  {Gukov}(2004)}]{Acharya:2004qe}%
  \BibitemOpen
  \bibfield  {author} {\bibinfo {author} {\bibfnamefont {B.~S.}\ \bibnamefont
  {Acharya}}\ and\ \bibinfo {author} {\bibfnamefont {S.}~\bibnamefont
  {Gukov}},\ }\href {\doibase 10.1016/j.physrep.2003.10.017} {\bibfield
  {journal} {\bibinfo  {journal} {Phys.Rept.}\ }\textbf {\bibinfo {volume}
  {392}},\ \bibinfo {pages} {121} (\bibinfo {year} {2004})},\ \Eprint
  {http://arxiv.org/abs/hep-th/0409191} {arXiv:hep-th/0409191 [hep-th]}
  \BibitemShut {NoStop}%
\bibitem [{\citenamefont {Acharya}\ and\ \citenamefont
  {Bobkov}(2010)}]{Acharya:2008hi}%
  \BibitemOpen
  \bibfield  {author} {\bibinfo {author} {\bibfnamefont {B.~S.}\ \bibnamefont
  {Acharya}}\ and\ \bibinfo {author} {\bibfnamefont {K.}~\bibnamefont
  {Bobkov}},\ }\href {\doibase 10.1007/JHEP09(2010)001} {\bibfield  {journal}
  {\bibinfo  {journal} {JHEP}\ }\textbf {\bibinfo {volume} {1009}},\ \bibinfo
  {pages} {001} (\bibinfo {year} {2010})},\ \Eprint
  {http://arxiv.org/abs/0810.3285} {arXiv:0810.3285 [hep-th]} \BibitemShut
  {NoStop}%
\bibitem [{\citenamefont {Giudice}\ and\ \citenamefont
  {Masiero}(1988)}]{Giudice:1988yz}%
  \BibitemOpen
  \bibfield  {author} {\bibinfo {author} {\bibfnamefont {G.}~\bibnamefont
  {Giudice}}\ and\ \bibinfo {author} {\bibfnamefont {A.}~\bibnamefont
  {Masiero}},\ }\href {\doibase 10.1016/0370-2693(88)91613-9} {\bibfield
  {journal} {\bibinfo  {journal} {Phys.Lett.}\ }\textbf {\bibinfo {volume}
  {B206}},\ \bibinfo {pages} {480} (\bibinfo {year} {1988})}\BibitemShut
  {NoStop}%
\bibitem [{\citenamefont {Brignole}\ \emph {et~al.}(2010)\citenamefont
  {Brignole}, \citenamefont {Ibanez},\ and\ \citenamefont
  {Munoz}}]{Brignole:1997dp}%
  \BibitemOpen
  \bibfield  {author} {\bibinfo {author} {\bibfnamefont {A.}~\bibnamefont
  {Brignole}}, \bibinfo {author} {\bibfnamefont {L.~E.}\ \bibnamefont
  {Ibanez}}, \ and\ \bibinfo {author} {\bibfnamefont {C.}~\bibnamefont
  {Munoz}},\ }\href {\doibase 10.1142/9789814307505_0004} {\bibfield  {journal}
  {\bibinfo  {journal} {Adv. Ser. Direct. High Energy Phys.}\ }\textbf
  {\bibinfo {volume} {21}},\ \bibinfo {pages} {244} (\bibinfo {year} {2010})},\
  \Eprint {http://arxiv.org/abs/hep-ph/9707209} {arXiv:hep-ph/9707209 [hep-ph]}
  \BibitemShut {NoStop}%
\bibitem [{\citenamefont {Dvali}(1996)}]{Dvali:1995hp}%
  \BibitemOpen
  \bibfield  {author} {\bibinfo {author} {\bibfnamefont {G.}~\bibnamefont
  {Dvali}},\ }\href {\doibase 10.1016/0370-2693(96)00022-6} {\bibfield
  {journal} {\bibinfo  {journal} {Phys.Lett.}\ }\textbf {\bibinfo {volume}
  {B372}},\ \bibinfo {pages} {113} (\bibinfo {year} {1996})},\ \Eprint
  {http://arxiv.org/abs/hep-ph/9511237} {arXiv:hep-ph/9511237 [hep-ph]}
  \BibitemShut {NoStop}%
\bibitem [{\citenamefont {Kilian}\ and\ \citenamefont
  {Reuter}(2006)}]{Kilian:2006hh}%
  \BibitemOpen
  \bibfield  {author} {\bibinfo {author} {\bibfnamefont {W.}~\bibnamefont
  {Kilian}}\ and\ \bibinfo {author} {\bibfnamefont {J.}~\bibnamefont
  {Reuter}},\ }\href {\doibase 10.1016/j.physletb.2006.09.033} {\bibfield
  {journal} {\bibinfo  {journal} {Phys.Lett.}\ }\textbf {\bibinfo {volume}
  {B642}},\ \bibinfo {pages} {81} (\bibinfo {year} {2006})},\ \Eprint
  {http://arxiv.org/abs/hep-ph/0606277} {arXiv:hep-ph/0606277 [hep-ph]}
  \BibitemShut {NoStop}%
\bibitem [{\citenamefont {Reuter}(2007)}]{Reuter:2007eh}%
  \BibitemOpen
  \bibfield  {author} {\bibinfo {author} {\bibfnamefont {J.}~\bibnamefont
  {Reuter}},\ }\href@noop {} {\  (\bibinfo {year} {2007})},\ \Eprint
  {http://arxiv.org/abs/0709.4202} {arXiv:0709.4202 [hep-ph]} \BibitemShut
  {NoStop}%
\bibitem [{\citenamefont {Howl}\ and\ \citenamefont
  {King}(2008)}]{Howl:2007zi}%
  \BibitemOpen
  \bibfield  {author} {\bibinfo {author} {\bibfnamefont {R.}~\bibnamefont
  {Howl}}\ and\ \bibinfo {author} {\bibfnamefont {S.}~\bibnamefont {King}},\
  }\href {\doibase 10.1088/1126-6708/2008/01/030} {\bibfield  {journal}
  {\bibinfo  {journal} {JHEP}\ }\textbf {\bibinfo {volume} {0801}},\ \bibinfo
  {pages} {030} (\bibinfo {year} {2008})},\ \Eprint
  {http://arxiv.org/abs/0708.1451} {arXiv:0708.1451 [hep-ph]} \BibitemShut
  {NoStop}%
\bibitem [{\citenamefont {Kolda}\ and\ \citenamefont
  {Martin}(1996)}]{Kolda:1995iw}%
  \BibitemOpen
  \bibfield  {author} {\bibinfo {author} {\bibfnamefont {C.~F.}\ \bibnamefont
  {Kolda}}\ and\ \bibinfo {author} {\bibfnamefont {S.~P.}\ \bibnamefont
  {Martin}},\ }\href {\doibase 10.1103/PhysRevD.53.3871} {\bibfield  {journal}
  {\bibinfo  {journal} {Phys. Rev.}\ }\textbf {\bibinfo {volume} {D53}},\
  \bibinfo {pages} {3871} (\bibinfo {year} {1996})},\ \Eprint
  {http://arxiv.org/abs/hep-ph/9503445} {arXiv:hep-ph/9503445 [hep-ph]}
  \BibitemShut {NoStop}%
\bibitem [{\citenamefont {Araki}\ \emph {et~al.}(2008)\citenamefont {Araki},
  \citenamefont {Kobayashi}, \citenamefont {Kubo}, \citenamefont
  {Ramos-Sanchez}, \citenamefont {Ratz} \emph {et~al.}}]{Araki:2008ek}%
  \BibitemOpen
  \bibfield  {author} {\bibinfo {author} {\bibfnamefont {T.}~\bibnamefont
  {Araki}}, \bibinfo {author} {\bibfnamefont {T.}~\bibnamefont {Kobayashi}},
  \bibinfo {author} {\bibfnamefont {J.}~\bibnamefont {Kubo}}, \bibinfo {author}
  {\bibfnamefont {S.}~\bibnamefont {Ramos-Sanchez}}, \bibinfo {author}
  {\bibfnamefont {M.}~\bibnamefont {Ratz}},  \emph {et~al.},\ }\href {\doibase
  10.1016/j.nuclphysb.2008.07.005} {\bibfield  {journal} {\bibinfo  {journal}
  {Nucl.Phys.}\ }\textbf {\bibinfo {volume} {B805}},\ \bibinfo {pages} {124}
  (\bibinfo {year} {2008})},\ \Eprint {http://arxiv.org/abs/0805.0207}
  {arXiv:0805.0207 [hep-th]} \BibitemShut {NoStop}%
\bibitem [{\citenamefont {Acharya}\ \emph {et~al.}(2014)\citenamefont
  {Acharya}, \citenamefont {Kane}, \citenamefont {Kumar}, \citenamefont {Lu},\
  and\ \citenamefont {Zheng}}]{Acharya:2014vha}%
  \BibitemOpen
  \bibfield  {author} {\bibinfo {author} {\bibfnamefont {B.~S.}\ \bibnamefont
  {Acharya}}, \bibinfo {author} {\bibfnamefont {G.~L.}\ \bibnamefont {Kane}},
  \bibinfo {author} {\bibfnamefont {P.}~\bibnamefont {Kumar}}, \bibinfo
  {author} {\bibfnamefont {R.}~\bibnamefont {Lu}}, \ and\ \bibinfo {author}
  {\bibfnamefont {B.}~\bibnamefont {Zheng}},\ }\href {\doibase
  10.1007/JHEP10(2014)001} {\bibfield  {journal} {\bibinfo  {journal} {JHEP}\
  }\textbf {\bibinfo {volume} {1410}},\ \bibinfo {pages} {1} (\bibinfo {year}
  {2014})},\ \Eprint {http://arxiv.org/abs/1403.4948} {arXiv:1403.4948
  [hep-ph]} \BibitemShut {NoStop}%
\bibitem [{\citenamefont {Banks}\ \emph {et~al.}(1995)\citenamefont {Banks},
  \citenamefont {Grossman}, \citenamefont {Nardi},\ and\ \citenamefont
  {Nir}}]{Banks:1995by}%
  \BibitemOpen
  \bibfield  {author} {\bibinfo {author} {\bibfnamefont {T.}~\bibnamefont
  {Banks}}, \bibinfo {author} {\bibfnamefont {Y.}~\bibnamefont {Grossman}},
  \bibinfo {author} {\bibfnamefont {E.}~\bibnamefont {Nardi}}, \ and\ \bibinfo
  {author} {\bibfnamefont {Y.}~\bibnamefont {Nir}},\ }\href {\doibase
  10.1103/PhysRevD.52.5319} {\bibfield  {journal} {\bibinfo  {journal}
  {Phys.Rev.}\ }\textbf {\bibinfo {volume} {D52}},\ \bibinfo {pages} {5319}
  (\bibinfo {year} {1995})},\ \Eprint {http://arxiv.org/abs/hep-ph/9505248}
  {arXiv:hep-ph/9505248 [hep-ph]} \BibitemShut {NoStop}%
\bibitem [{\citenamefont {Acharya}\ \emph {et~al.}(2011)\citenamefont
  {Acharya}, \citenamefont {Kane}, \citenamefont {Kuflik},\ and\ \citenamefont
  {Lu}}]{Acharya:2011te}%
  \BibitemOpen
  \bibfield  {author} {\bibinfo {author} {\bibfnamefont {B.~S.}\ \bibnamefont
  {Acharya}}, \bibinfo {author} {\bibfnamefont {G.}~\bibnamefont {Kane}},
  \bibinfo {author} {\bibfnamefont {E.}~\bibnamefont {Kuflik}}, \ and\ \bibinfo
  {author} {\bibfnamefont {R.}~\bibnamefont {Lu}},\ }\href {\doibase
  10.1007/JHEP05(2011)033} {\bibfield  {journal} {\bibinfo  {journal} {JHEP}\
  }\textbf {\bibinfo {volume} {1105}},\ \bibinfo {pages} {033} (\bibinfo {year}
  {2011})},\ \Eprint {http://arxiv.org/abs/1102.0556} {arXiv:1102.0556
  [hep-ph]} \BibitemShut {NoStop}%
\bibitem [{\citenamefont {Dreiner}(2010)}]{Dreiner:1997uz}%
  \BibitemOpen
  \bibfield  {author} {\bibinfo {author} {\bibfnamefont {H.~K.}\ \bibnamefont
  {Dreiner}},\ }\href {\doibase 10.1142/9789814307505_0017} {\bibfield
  {journal} {\bibinfo  {journal} {Adv.Ser.Direct.High Energy Phys.}\ }\textbf
  {\bibinfo {volume} {21}},\ \bibinfo {pages} {565} (\bibinfo {year} {2010})},\
  \Eprint {http://arxiv.org/abs/hep-ph/9707435} {arXiv:hep-ph/9707435 [hep-ph]}
  \BibitemShut {NoStop}%
\bibitem [{\citenamefont {Svrcek}\ and\ \citenamefont
  {Witten}(2006)}]{Svrcek:2006yi}%
  \BibitemOpen
  \bibfield  {author} {\bibinfo {author} {\bibfnamefont {P.}~\bibnamefont
  {Svrcek}}\ and\ \bibinfo {author} {\bibfnamefont {E.}~\bibnamefont
  {Witten}},\ }\href {\doibase 10.1088/1126-6708/2006/06/051} {\bibfield
  {journal} {\bibinfo  {journal} {JHEP}\ }\textbf {\bibinfo {volume} {0606}},\
  \bibinfo {pages} {051} (\bibinfo {year} {2006})},\ \Eprint
  {http://arxiv.org/abs/hep-th/0605206} {arXiv:hep-th/0605206 [hep-th]}
  \BibitemShut {NoStop}%
\bibitem [{\citenamefont {Acharya}\ \emph {et~al.}(2010)\citenamefont
  {Acharya}, \citenamefont {Bobkov},\ and\ \citenamefont
  {Kumar}}]{Acharya:2010zx}%
  \BibitemOpen
  \bibfield  {author} {\bibinfo {author} {\bibfnamefont {B.~S.}\ \bibnamefont
  {Acharya}}, \bibinfo {author} {\bibfnamefont {K.}~\bibnamefont {Bobkov}}, \
  and\ \bibinfo {author} {\bibfnamefont {P.}~\bibnamefont {Kumar}},\ }\href
  {\doibase 10.1007/JHEP11(2010)105} {\bibfield  {journal} {\bibinfo  {journal}
  {JHEP}\ }\textbf {\bibinfo {volume} {1011}},\ \bibinfo {pages} {105}
  (\bibinfo {year} {2010})},\ \Eprint {http://arxiv.org/abs/1004.5138}
  {arXiv:1004.5138 [hep-th]} \BibitemShut {NoStop}%
\bibitem [{\citenamefont {Joyce}(2000)}]{G2}%
  \BibitemOpen
  \bibfield  {author} {\bibinfo {author} {\bibfnamefont {D.}~\bibnamefont
  {Joyce}},\ }\href@noop {} {\emph {\bibinfo {title} {Compact manifolds with
  special holonomy}}}\ (\bibinfo  {publisher} {Oxford Mathematical Monographs,
  Oxford University Press, Oxford},\ \bibinfo {year} {2000})\BibitemShut
  {NoStop}%
\bibitem [{\citenamefont {{Corti}}\ \emph {et~al.}(2012)\citenamefont
  {{Corti}}, \citenamefont {{Haskins}}, \citenamefont {{Nordstr{\"o}m}},\ and\
  \citenamefont {{Pacini}}}]{2012arXiv1207.4470C}%
  \BibitemOpen
  \bibfield  {author} {\bibinfo {author} {\bibfnamefont {A.}~\bibnamefont
  {{Corti}}}, \bibinfo {author} {\bibfnamefont {M.}~\bibnamefont {{Haskins}}},
  \bibinfo {author} {\bibfnamefont {J.}~\bibnamefont {{Nordstr{\"o}m}}}, \ and\
  \bibinfo {author} {\bibfnamefont {T.}~\bibnamefont {{Pacini}}},\ }\href@noop
  {} {\bibfield  {journal} {\bibinfo  {journal} {ArXiv e-prints}\ } (\bibinfo
  {year} {2012})},\ \Eprint {http://arxiv.org/abs/1207.4470} {arXiv:1207.4470
  [math.DG]} \BibitemShut {NoStop}%
\end{thebibliography}%

\end{document}